\documentclass[preprints,article,accept,moreauthors,pdftex]{mdpi}
\firstpage{1}
\pubvolume{1}
\issuenum{1}
\articlenumber{0}
\pubyear{2021}
\copyrightyear{2021}
\datereceived{}
\dateaccepted{}
\datepublished{}
\hreflink{https://doi.org/} % If needed use \linebreak

\Title{On-axis Optical Bench for Laser Ranging Instruments in future gravity missions}

\TitleCitation{On-axis Optical Bench for Laser Ranging Instruments in future gravity missions}

\Author{Yichao Yang $^{1,3}$\orcidA{}, Kohei Yamamoto $^{2}$\orcidB{}, Miguel Dovale \'Alvarez  $^{2}$*\orcidC{}, Daikang Wei  $^{2}$\orcidD{}, Juan Jos\'e Esteban Delgado $^{2}$\orcidE{}, Vitali M\"{u}ller $^{2}$, Jianjun Jia $^{1,4}$ and Gerhard Heinzel $^{2}$*\orcidF{}}

\AuthorNames{Yichao Yang, Kohei Yamamoto, Miguel Dovale \'Alvarez, Juan Jose Esteban Delgado, Vitali M\"{u}ller, Jianjun Jia and Gerhard Heinzel}

\AuthorCitation{Yang, Y.; Yamamoto, K.; Dovale \'Alvarez, M.; Esteban Delgado, J.J.; Heinzel,G.; M\"{u}ller,V.; Jia, J.}

\address{%
$^{1}$ \quad Key Laboratory of Space Active Opto-electronics Technology, Shanghai Institute of Technical Physics, Chinese Academy of Sciences, Shanghai 200083, China; yangyichao@mail.sitp.ac.cn\\
$^{2}$ \quad Max-Planck-Institut f\"ur Gravitationsphysik (Albert-Einstein-Institut) and Institut f\"ur Gravitationsphysik, Leibniz Universit\"at Hannover, Callinstrasse 38, D-30167 Hannover, Germany; kohei.yamamoto@aei.mpg.de (K.Y.); miguel.dovale@aei.mpg.de (M.D.A.); juan.jose.esteban@aei.mpg.de (J.J.E.D.); gerhard.heinzel@aei.mpg.de (G.H.)\\
$^{3}$ \quad School of Electronic, Electrical and Communication Engineering, University of Chinese Academy of Sciences, Beijing 100049, China;\\
$^{4}$ \quad School of Physical Sciences, University of Chinese Academy of Sciences, Beijing 100049, China;}
\corres{Correspondence: miguel.dovale@aei.mpg.de (M.D.A.) and gerhard.heinzel@aei.mpg.de (G.H.)}

\abstract{The Laser Ranging Interferometer onboard the Gravity Recovery and Climate Experiment Follow-On mission proved the feasibility of an interferometric sensor for inter-satellite length tracking with sub-nanometer precision, establishing an important milestone for space laser interferometry and the general expectation that future gravity missions will employ heterodyne laser interferometry for satellite-to-satellite ranging. In this paper we present the design of an on-axis optical bench for next-generation laser ranging which enhances the received optical power and the transmit beam divergence, enabling longer interferometer arms and relaxing the optical power requirement of the laser assembly. All design functionalities and requirements are verified by means of computer simulations. A thermal analysis is carried out to investigate the robustness of the proposed optical bench to the temperature fluctuations found in orbit.}


\begin{document}

\section{Introduction}

The Gravity Recovery and Climate Experiment (GRACE) mission~\cite{Tapley2004}, in orbit since March 2002 to the end of its science mission in October 2017, was a joint mission of the National Aeronautics and Space Administration (NASA) and the German Aerospace Center (DLR). Twin satellites, in a trailing formation flying $\sim 500$\,km above the Earth with a nominal separation of 200\,km on a near polar orbit, mapped the variations of the Earth's gravity field. The primary science instrument was an inter-satellite microwave ranging system in K/Ka-band that tracked the distance between the satellites and its rate of change with a sensitivity down to 0.1\,$\upmu$m/s, enabling the detection of the tiny variations in the Earth's gravitational pull over the trailing satellites. To continue observing the Earth's varying gravity field, the Gravity Recovery and Climate Experiment Follow-On (GRACE-FO) mission~\cite{Kornfeld2019} was launched in May 2018, again as a US-Germany collaboration.

A key feature of the GRACE-FO mission was the inclusion of the Laser Ranging Interferometer (LRI) as a technology demonstrator to perform the range measurement with higher precision than its microwave counterpart. Using the range measurement data in combination with precise orbit determination via Global Positioning System (GPS) instruments and on-board accelerometers, scientists regularly construct a detailed monthly gravity field map of the Earth, providing a unique global view of the Earth's surface mass distribution.

The GRACE and GRACE-FO observations reveal the effects of climate change over time, and enable future predictions, quantifying the loss of mass from ice sheets and glaciers, the contribution of water influx to sea level rise, as well as monitoring changes in underground water reserves, and in the hydrological cycle~\cite{Tapley2019, Velicogna2020, Landerer2020}.

The LRI in GRACE-FO enabled the first laser interferometer link between distant satellites, and provided range measurements with a sensitivity of 200\,pm/$\sqrt{\text{Hz}}$ at 1\,Hz, more than 3 orders of magnitude lower than the $0.6\,\upmu\mathrm{m}/\sqrt{\mathrm{Hz}}$ noise of the primary microwave ranging instrument~\cite{Spero2021}. The GRACE-FO LRI also served as a technology demonstrator for the ESA-NASA Laser Interferometer Space Antenna (LISA) mission~\cite{LISA}, as well as the Chinese space-borne gravitational wave observatories Taiji~\cite{Luo2020} and TianQin~\cite{Luo2016} missions. 

The GRACE-FO LRI, as a technology demonstrator and not the primary science instrument, had restrictions on its design, e.g., volume, mass, instrument accommodation, and relaxed requirements on reliability and lifetime. After the success of the GRACE-FO LRI, the next generation of gravity missions are expected to use laser interferometry as the primary measurement principle.

The optical bench (OB) is a central part of the LRI. Its primary purpose is to receive light from the remote satellite and combine it with a locally generated reference beam (LO) in order to extract their relative phase. The most promising phase measurement principle continues to be optical heterodyne detection~\cite{Mandel2020, Muller2017PhD, Francis2017}. In addition, the OB emits laser light with a well-defined phase that is obtained by phase-locking to the received light. A secondary purpose of the OB is to measure the angle of the incoming light w.r.t.~an OB frame of reference. This allows inference of the misalignment of the receiver S/C relative to the inter-satellite line-of-sight. This information is used in the LRI to actuate on the pointing of the transmitted beam to the remote S/C, and can be used in future missions to actuate in the attitude of the local S/C, in order to enhance the optical link power, keep the link stable, and minimize secondary noise sources.

\begin{figure}
\centering
\includegraphics[scale=0.8]{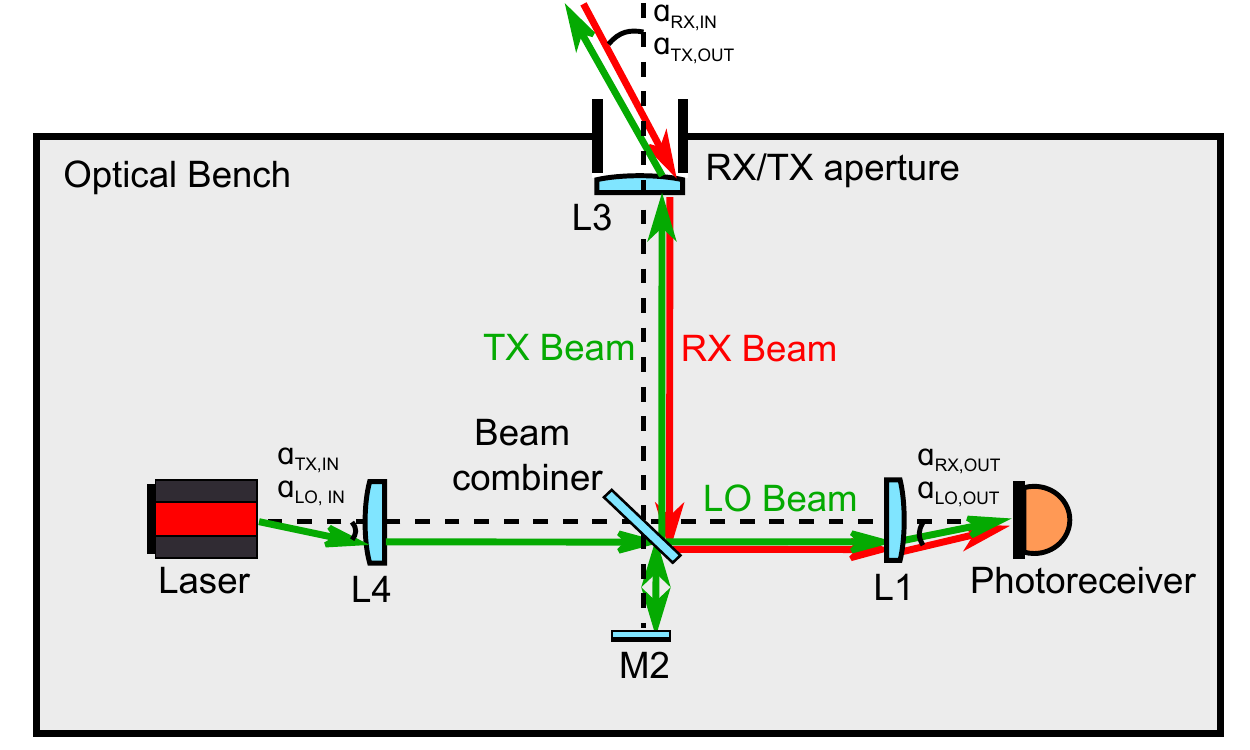}
\caption{Design concept of an on-axis optical bench for a next generation LRI. The receive and transmit beams (RX and TX respectively) propagate into and out of the optical bench through the same aperture, where their paths coincide.}
\label{figure:on-axis}
\end{figure}

In the GRACE-FO mission, the LRI OB followed an ``off-axis'' design~\cite{Nicklaus2017}, where the receiving (RX) light and the transmitting (TX) light propagate into and out of the OB through separate apertures, namely the RX and TX apertures, and follow different paths along the OB. This type of design can lead to instruments with fewer optical components compared to the ``on-axis'' topology, where the RX and TX beam paths coincide, and the OB features a single aperture (Figure~\ref{figure:on-axis}). The on-axis topology requires using polarizing optics in order to optimally combine the RX and LO light at the photodetectors, potentially yielding a more complex optical setup. However, due to the shared optical path between the RX and TX beams, an on-axis design has the potential to use a single baffle and a single telescope for both RX light reception and TX light transmission. The OB design based on a lens retroreflector proposed by M\"uller~\cite{Muller2017PhD}, and the design based on a corner cube retroreflector proposed by Mandel et al.~\cite{Mandel2020}, are both examples of on-axis topologies. For optical benches of space-borne interferometers with baselines longer than $10^6$ meters, such as those proposed for gravitational wave detection missions, the on-axis OB topology is the scheme of choice.

In this paper we present the design of an on-axis OB for laser ranging in future gravity missions. Our design presents the following features:
\begin{itemize}
	\item Two quadrant photoreceivers in balanced detection configuration capture the length and angular signals of the RX-LO heterodyne interference.
	\item Five two-lens imaging systems using a total of four lenses (each lens is shared by more than one system) serve as beam compressors for the LO and RX beams, and beam expander for the TX beam, enhancing the light collecting area and decreasing the TX beam far field divergence;
	\item A fast steering mirror actuates on the tip and tilt of the LO beam using the differential wavefront sensing signals of the RX-LO beatnote, maximizing the heterodyne efficiency of the interferometer, thus yielding phase signals with optimal signal-to-noise ratio, regardless of the attitude of the local S/C;
	\item The RX and LO imaging systems minimize the RX and LO beam walk at the photodiodes irrespective of the attitude of the local satellite, greatly reducing the coupling of attitude jitter to the inter-satellite range measurement, and keeping the detection system in a stable operating point;
    \item The TX imaging system, in combination with the actuation of the fast steering mirror on the locally generated beam, ensures accurate TX beam pointing to the remote S/C, eliminating the need for a retroreflector;
    \item The imaging system magnifications can be tuned to flexibly adjust the size of the beams to match the RX/TX aperture, as well as the active area of the photodiodes.
\end{itemize}

The paper is structured as follows: In Section~\ref{section:ob_design} we detail the design of the proposed OB. In Section~\ref{section:results} we present the computer model of the optical bench and the results of simulating S/C angular motion, including an account of thermoelastic effects and refractive index variations due to in-orbit temperature fluctuations. Finally, in Section~\ref{section:conclusions} we summarize our findings.

\section{On-axis LRI optical bench design}
\label{section:ob_design}

Some general design requirements apply to the OB regardless of the topology. One key requirement is that a sufficient amount of light is transferred from the receive aperture (RX aperture) to the photodetectors in order to enable stable phase-tracking whilst maintaining the phase fidelity of the light by mitigating disturbances from diffraction due to clipping and from stray light. For a wavelength of approximately 1\,$\upmu$m the received light power should exceed 10\,pW at the photoreceivers~\cite{Muller2017PhD}, since that corresponds to a carrier-to-noise density (CNR) of $\sim 70$\,dB-Hz considering the current technology of photoreceivers, lasers and phasemeters. Though phase-tracking at lower CNR values has been demonstrated in lab experiments~\cite{Francis2014, Dick2008}, a value of 70\,dB-Hz guarantees cycle-slip-free tracking with sufficient margin and robustness against disturbances from thruster firings, sun-blindings, micro-meteorite impacts, etc. An automatic beam alignment system using a fast steering mirror (FSM) can further enhance the robustness, since it maximizes the interferometric contrast and therefore the CNR with high gain and bandwidth at the local S/C, and ideally also the transmitted power to the remote S/C due to an optimization of the direction of the transmitted beam.

Another key requirement of the OB is the feature of a well-defined and stable virtual reference point (RP) that is physically accessible in order for it to be co-located with the S/C center-of-mass (c.m.), such that the measured range is invariant under small rotations of the S/C around this point in an arbitrary axis. Considering integration tolerances, thermal variations, launch vibrations, thermo-elastic deformations and settling effects from the transition to space, a reasonable value for the in-flight tilt-to-length (TTL) coupling is of the order of $100\,\upmu$m/rad, which has been demonstrated by the GRACE-FO LRI \cite{Wegener2020}, and, to some extent, can be reduced further by in-flight calibrations.

The OB also has to ensure that the RX and LO beams remain well aligned, maintaining good interferometric contrast regardless of the attitude of the local S/C. Moreover, the OB has to provide a TX beam that is well aligned to the inter-satellite line-of-sight, thereby optimizing the optimal link. The acceptable pointing stability is typically of the order of the divergence angle. For laser interferometers employing fundamental Gaussian beams, the intensity at the misalignment angle drops by $e^{-2}=13.5$\%, or by 8.6\,dB in terms of CNR. Typical divergence angles are of the order of $100\,\upmu$rad (see Table~\ref{table:configurations} for exact values of the TX beam divergence in various optical bench configurations).

\begin{figure}
    \centering
    \includegraphics[scale=0.2]{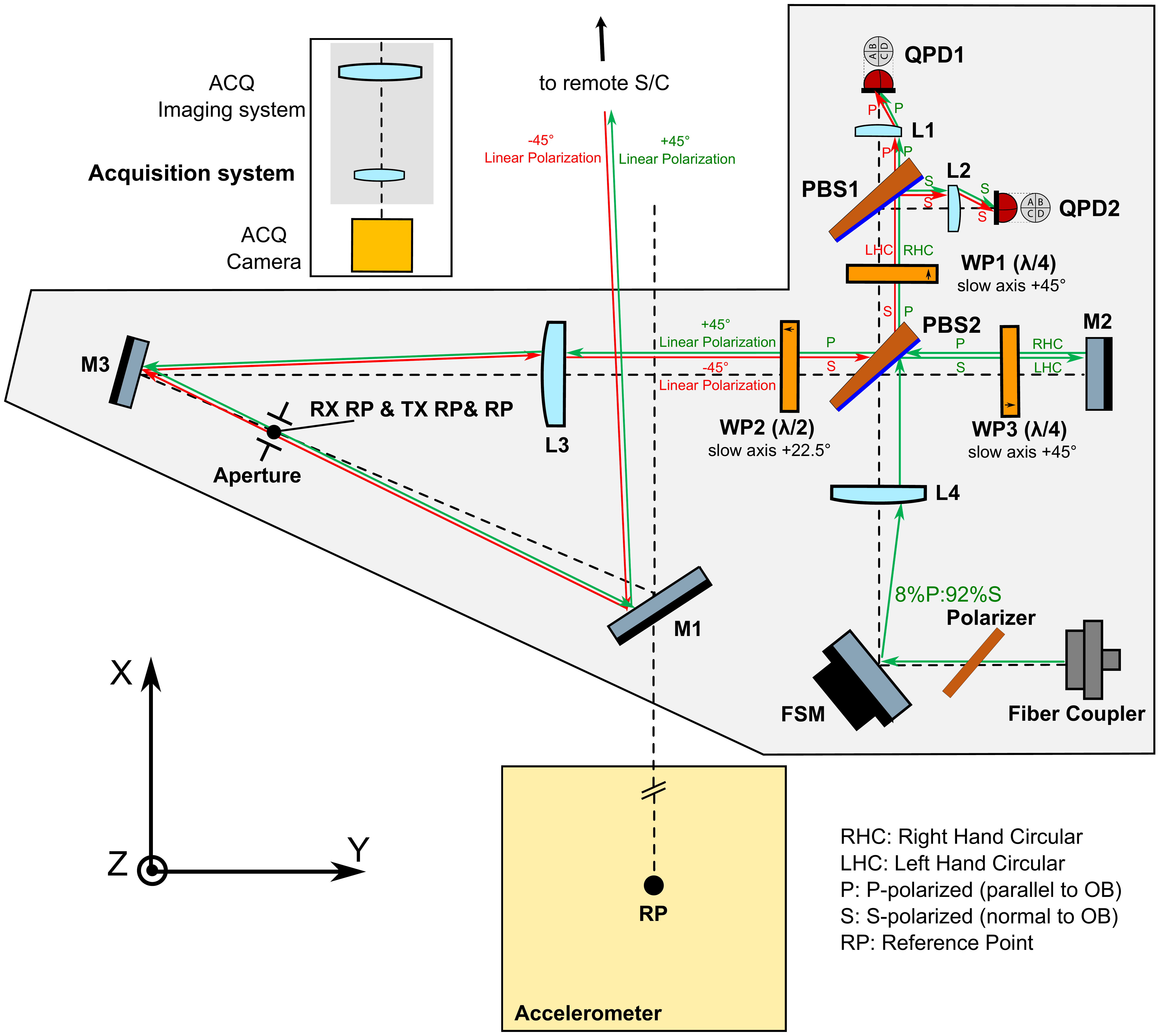}
    \caption{Optical bench layout. The green arrows depict the path and direction of the local oscillator and transmit beams, which originate from a single beam injected into the bench via a fiber injector. The red arrows depict the path and direction of the receive beam, which couples into the bench via mirror M1. The polarization state of each beam is indicated (RHC: right hand circular; LHC: left hand circular). The receive and transmit reference points (RX/TX RP) coincide in the left focal plane of lens L3, along the optical axis between mirrors M3 and M1. The RX/TX RP is imaged into the spacecraft center-of-mass, where an accelerometer is located. The range measurement is invariant under rotations of the spacecraft around this point. The receive and local oscillator beams interfere at polarizing beamsplitter PBS1, and are captured by the pair of quadrant photodiodes QPD1 and QPD2 in balanced detection configuration.}
    \label{figure:setup}	
\end{figure}

Based on the aforementioned requirements, an optical bench design has been carried out as part of the development of a laser ranging instrument for future gravity missions. The design comprises a series of Keplerian telescope imaging systems, in contrast with other realizations employing corner cube retroreflectors, such as that proposed by Mandel et al~\cite{Mandel2020} and by Nicklaus et al~\cite{Nicklaus2020}.

TTL coupling is the leading source of noise in the range measurement at low frequencies. Careful optimization of the optical bench leads to a design TTL coupling well below 100$\,\upmu$m/rad. However, it is expected that the final in-orbit TTL coupling is larger, at the level of $\sim 100\,\upmu$m/rad, due to integration tolerances, settling effects from ground to space, and uncertainties and variations in the c.m.\ position. These external effects make a prediction of the low-frequency ranging sensitivity difficult, but future missions are expected to exhibit lower satellite pointing errors that drive the TTL error~\cite{Nicklaus2020} and post-processing correction of the TTL could further reduce the dominant TTL error~\cite{Wegener2020}.

At higher frequencies (above $\sim$40\,mHz) the sensitivity is limited by residual frequency fluctuations of the laser, arising from the instability of a reference optical cavity or of an atomic standard. The coupling of laser frequency noise into the range measurement can be estimated by the fractional frequency noise of the laser scaled by the S/C separation~\cite{Sheard2012}. For example, the 1064\,nm laser on GRACE-FO was stabilized in-flight to a frequency noise below 1\,Hz/$\sqrt{\mathrm{Hz}}$ resulting in a ranging sensitivity below 1\,nm/$\sqrt{\mathrm{Hz}}$ for a S/C separation of approximately 200\,km and at a Fourier frequency of 1\,Hz~\cite{Abich2019}. At large inter-satellite distances, the interferometer may become shot-noise-limited due to the decreased received optical power, resulting in a CNR-limited sensitivity. Optical bench designs with intrinsically higher CNR values can therefore lead to a sensitivity enhancement and enable longer baselines.

The design of the proposed OB is depicted in Figure~\ref{figure:setup}. The RX beam enters the OB via mirror M1, which images the virtual RP (i.e., the S/C c.m.) into the RX/TX reference point aperture located in the OB. The LO and TX beams both originate from a single beam that is injected into the bench via a fiber injector. After passing a suitably aligned polarizer, this initial beam contains 8\% p-polarized and 92\% s-polarized light, and through interacting with the polarizing beamsplitter PBS2, the beam is split into the TX beam (p-pol), which is transmitted out of the spacecraft via the same path as the RX beam, and the LO beam (s-pol), which propagates to PBS1 where it interferes with the RX beam. The RX-LO beatnote is then captured by both quadrant photodiodes QPD1 and QPD2 in a balanced detection scheme. A total of three retarder waveplates (WP1, WP2, and WP3) are placed north, west and east of PBS2 in order to keep all beams in orthogonal polarization states prior to the RX-LO interference at PBS1. The main purpose of introducing WP2 with its slow axis at 22.5 degrees is to allow both the local and the remote S/C optical benches to share identical optical layouts, as they both receive and transmit beams with -45 and +45 degree linear polarizations respectively.

Mirrors M3 and M1 fold the light path, reducing the footprint of the OB, and providing sufficient space for the installation of the accelerometer, also depicted in Figure~\ref{figure:setup}. Mirror M1 can also deflect the light out of the XY plane, in case that the S/C c.m. were out-of-plane from the OB. The position and orientation of M1 can be adjusted during the integration of the instrument to achieve collocation of the S/C c.m. and the RP, which greatly simplifies the placement of the OB itself.

The acquisition system is composed of an imaging system and a focal plane detector to image the RX light. Such a scheme is commonly used in space laser communication terminals~\cite{Kaushal2017}. A critical requirement is the alignment stability of the optical axis of the interferometer with respect to the acquisition system, which is in our opinion a well-known and solvable challenge for space missions with optical instruments.

Angular motion of the local spacecraft perturbs the RX-LO wavefront overlap, yielding average and huge differential and parasitic common-mode phase changes at the different segments of the quadrant photodiodes. In contrast, angular motion of the distant spacecraft yields changes of the received light power, but leaves the wavefront overlap largely unaffected. The differential phase between QPD segments, i.e., the differential wavefront sensing signals $\text{DWS}_{\text{h}}$ and $\text{DWS}_{\text{v}}$~\cite{Heinzel2020}, measure the relative tilt and tip between the interfering fields. In consequence, the DWS signals can be used in a feedback loop for transmit beam pointing or for spacecraft attitude control. In our OB design, two independent control loops are used to actuate the tip and tilt degrees of the FSM, yielding $\text{DWS}_{\text{h}} \sim 0$ and $\text{DWS}_{\text{v}} \sim 0$, as sensed by one of the quadrant photoreceivers, thus ensuring that the interfering fields' phasefronts are aligned at the detectors regardless of the S/C attitude.

The cross coupling of the angular jitter of the spacecraft into the interferometric range measurement, or TTL coupling, is of critical importance, and can be suppressed to a large extent by means of employing suitable imaging systems (IS). These consist of a set of lenses that are placed in the beam path, and together image the point of rotation of the beam onto the center of the photodiodes such that the detector is in a pupil plane where beam walk is minimized while the angular tilt of the wavefront is maximized. They also serve the purpose of compressing or enlarging the beams according to what is desired, e.g., to adapt the size of the interfering beams to the active area of the photodiodes, or to enlarge the size of the transmit beam in order to decrease its divergence. In the proposed OB architecture, a total of five two-lens Keplerian telescope imaging systems are implemented: two of the so-called RX IS, which share lens L3 (L3-L1 and L3-L2); two of the so-called LO IS, which share lens L4 (L4-L1 and L4-L2); and the so-called TX IS (L4-L3). All of these imaging systems contain mirrors, beamsplitters and waveplates between the first and second lens.

The RX IS's keep the received light beam at a fixed position on the photoreceivers independent of the incidence angle at the receive aperture, and the LO IS's do the same for the local oscillator beam independent of the steering mirror state. On the other hand, the TX IS ensures that, when the FSM actuation with closed DWS loops is engaged, the transmit beam propagates antiparallel to the received light, i.e., the local spacecraft ``points'' accurately at the remote spacecraft with its TX beam.

In the RX and LO imaging systems, the first lens acts on only one of the two interfering beams, while the second lens acts on both beams. The resulting configuration departs significantly from other interferometers employing imaging systems to suppress TTL coupling, where all lenses are placed after the interferometer's beam combiner~\cite{Chwalla2020, Nicklaus2017, Muller2017PhD, Mandel2020}, and is therefore the subject of scrutiny in this paper.

 The lenses used in the model are available off-the-shelf from a popular lens manufacturer. The use of free-form lenses for TTL coupling reduction in similar precision interferometers has been studied in~\cite{Schuster17PhD}, and does not offer a significant advantage in terms of raw TTL coupling suppression. Fused silica substrates are chosen due to the excellent transmission properties and low coefficient of thermal expansion.

A notable feature of this setup is that a relationship is established between the angular magnifications of the RX, LO, and TX imaging systems,
\begin{equation}
\label{three_m_abs}
|m_{\text{RX}}|=\frac{|m_{\text{LO}}|}{|m_{\text{TX}}|},
\end{equation}
where $m_{\text{RX}}$, $m_{\text{TX}}$ and $m_{\text{LO}}$ are the angular magnifications of the respective imaging systems. The signs of the angular magnifications depend on the specific OB layout. The beam angle w.r.t.~the optical axis at the entrance pupil of each imaging system ($\alpha_{i,\text{IN}}$), and the angle at the exit pupil ($\alpha_{i,\text{OUT}}$) are given by (we work in two-dimensional space for simplicity)
\begin{equation}
\alpha_{i,\text{OUT}} = m_{i}\cdot\alpha_{i, \text{IN}},
\label{equation:input-output-angles}
\end{equation}
where $i=\left\{\text{RX,~LO,~TX}\right\}$. The sign of these angles follows the Cartesian Sign Convention: acute angles are positive when produced by anti-clockwise rotation from the optical axis, and negative when produced by clockwise rotation. In our design $\alpha_{\text{TX,IN}} = \alpha_{\text{LO,IN}}$, since the TX and LO imaging systems share lens L4 (see Figure~\ref{figure:setup}). Moreover, by actuating the FSM using DWS, the setup ensures that $\alpha_{\text{RX,OUT}}=\alpha_{\text{LO,OUT}}$. Therefore, assuming that $m_{\text{RX}}= m_{\text{LO}} / m_{\text{TX}}$ can be obtained through a specific OB layout, Equation~\ref{equation:input-output-angles} can be used to show that the setup can ideally produce perfect TX beam pointing to the remote S/C,
\begin{equation}
\label{RX_IN_and_TX_OUT}
\alpha_{RX,IN} = \frac{m_{LO}}{m_{RX}}\cdot\alpha_{LO,IN}=m_{TX}\cdot\alpha_{TX,IN}=\alpha_{TX,OUT}.
\end{equation}
The sign of the in-plane angular magnification of a Keplerian telescope is determined by the number of reflections in its optical path. An even (odd) number of reflections on the optical path leads to negative (positive) angular magnification. Using beam splitters and mirrors, the OB can therefore be designed such that $m_{\text{RX}}= m_{\text{LO}} / m_{\text{TX}}$ is verified.
   
To illustrate the flexibility of the proposed OB design, Table~\ref{table:configurations} shows the resulting heterodyne efficiency $\eta$ and carrier-to-noise density CNR for two OB configurations (``A'' and ``B'') differing in the radius of the injected local beam. Both configurations feature an aperture of radius $r_{\text{TX/RX AP}}=8\,$mm, and QPDs with active area radii of 0.5\,mm. The effective focal lengths of the selected lenses, measured by 589\,nm light, are: $f_{\text{L1}}=f_{\text{L2}}=12.7\,$mm, $f_{\text{L3}}=200\,$mm, and $f_{\text{L4}}=75.6\,$mm. The best performance out of the two proposed configurations is achieved in the ``A'' configuration, yielding a heterodyne efficiency of 0.85 and a CNR of 94\,dB.

The two aforementioned OB configurations are compared in Table~\ref{table:configurations} against the original GRACE-FO LRI OB, as well as two modified versions with the same aperture design and RX IS lenses as the new LRI OB ``A'' design. As illustrated by the ``Modified 1'' design, since the magnification of each beam cannot be adjusted independently in GRACE-FO, the beam size mismatch between the RX and LO beams at the surface of the photodetectors results in a degradation of heterodyne efficiency. Even if the RX and TX apertures are increased, the CNR does not improve significantly compared to the original design. As illustrated by the ``Modified 2'' design, greatly enlarging the waist radius of the local oscillator beam leads to similar CNR performance as the new LRI OB ``A'' design, at the expense of the increased optical complexity of achieving such a large waist size.
\begin{table}
\renewcommand{\arraystretch}{1.25}
\center
\caption{Comparison of some optical parameters between two configurations of the proposed LRI OB design, the GRACE-FO LRI OB, and two modified versions of the GRACE-FO LRI OB. $r_{\text{TX/RX AP}}$ is the receive/transmit aperture; $\omega_{0}$ is the waist radius of the local oscillator beam after the fiber injector; $m_i^{-1}$ with $i=\left\{\text{RX,~LO,~TX}\right\}$ are the linear magnifications of the RX, LO and TX imaging systems respectively; $\eta$ is the heterodyne efficiency of the RX-LO beatnote; $\theta_{\mathrm{TX}}$ is the half-angle divergence of the TX beam out of the aperture; CNR (ideal) is the resulting carrier-to-noise density for perfectly aligned local and remote satellites; CNR (misaligned) is the resulting carrier-to-noise density considering a $50\,\upmu$rad misalignment of the local S/C with respect to the line-of-sight to the remote S/C~\cite{Nicklaus2017,Muller2017PhD}.}
\label{table:configurations}
\begin{tabular}{r c c c c c}
\toprule
 & \rotatebox[origin=c]{80}{New LRI OB ``A''} & \rotatebox[origin=c]{80}{New LRI OB ``B''} & \rotatebox[origin=c]{80}{GRACE-FO} & \rotatebox[origin=c]{80}{GRACE-FO Mod 1} & \rotatebox[origin=c]{80}{GRACE-FO Mod 2} \\
\midrule
$r_{\text{TX/RX AP}}$ & 8\,mm & 8\,mm & 4\,mm & 8\,mm & 8\,mm \\
$\omega_{0}$ & 2.5\,mm & 1\,mm & 2.5\,mm & 2.5\,mm & 6.6\,mm \\
$|m_{\text{RX}}^{-1}|$ & 0.064 & 0.064 & 0.125 & 0.064 & 0.064 \\
$|m_{\text{LO}}^{-1}|$ & 0.168 & 0.168 & 0.125 & 0.064 & 0.064 \\
$|m_{\text{TX}}^{-1}|$ & 2.646 & 2.646 & n/a & n/a & n/a \\
$\eta$ & 0.85 & 0.23 & 0.67 & 0.2 & 0.85  \\
$\theta_{\mathrm{TX}}$ & 65\,$\upmu$rad & 131\,$\upmu$rad & 149\,$\upmu$rad & 138\,$\upmu$rad & 65\,$\upmu$rad  \\
CNR (ideal) & 94.2\,dB & 82.9\,dB & 81.2\,dB & 82.5\,dB & 94.9\,dB \\
CNR (misaligned) & 88.8\,dB & 81.4\,dB & 80.0\,dB & 81.2\,dB & 89.5\,dB \\
\bottomrule
\end{tabular}
\end{table}

\section{Optical simulations}
\label{section:results}

In order to realize the OB design, we built an optical model using the interferometer simulation software IFOCAD~\cite{Ifocad}, a collection of C++ libraries for simulating laser interferometers. The OB components, such as the beam splitters and mirrors, the lenses, and the photoreceivers, are parametrized and included in the model. IFOCAD provides methods for tracing general astigmatic Gaussian beams through three-dimensional space, as well as for computing the relevant two-beam interferometric signals, such as the heterodyne efficiency, the longitudinal pathlength signal (LPS) and differential wavefront sensing (DWS) signals~\cite{Wanner12OC}. The LPS sensed by a single element photodiode is computed as
\begin{equation}
	s_{\text{LPS}} \equiv \frac{\phi}{k} = s + \frac{1}{k}\arg\left(\iint  E_1 E_2 ^{\ast} dS_{\text{pd}} \right),
	\label{equation:LPS}
\end{equation}
where $\phi$ is the interferometric phase, $k=2\pi/\lambda$ is the wave number, $s$ is the macroscopic accumulated optical pathlength difference during beam propagation through the setup, $E_1$ and $E_2$ are the complex amplitudes of the interfering beams, and $dS_{\text{pd}}$ is a surface element in the detector surface. There exist several variants of the LPS signal for a QPD~\cite{Wanner2015}, of which we use the ``average phase'' definition, where the LPS is computed as the average phase over the four detector segments $s_{\text{LPS}} = (\phi_A+\phi_B+\phi_C+\phi_D)/4k$, where $\phi_{A\dots D}$ are the interferometric phases measured by each of the QPD segments. Note that $s_{\text{LPS}}$ takes into account any effects stemming from the transverse distributions of the two interfering beams in the detector plane, as well as Gouy phase shifts, which are known to be potential sources of error in precision interferometers~\cite{Yoshino2020}. Polarization and stray light effects are neglected in the simulation, and will be the subject of future investigations. Parasitic polarization states, polarization fluctuations, and ghost beams, can couple to the LPS and DWS signals, and could have an impact on the interferometer performance~\cite{Kaune2021PhD, Isleif2018PhD}.

\begin{figure}
\includegraphics[scale=1]{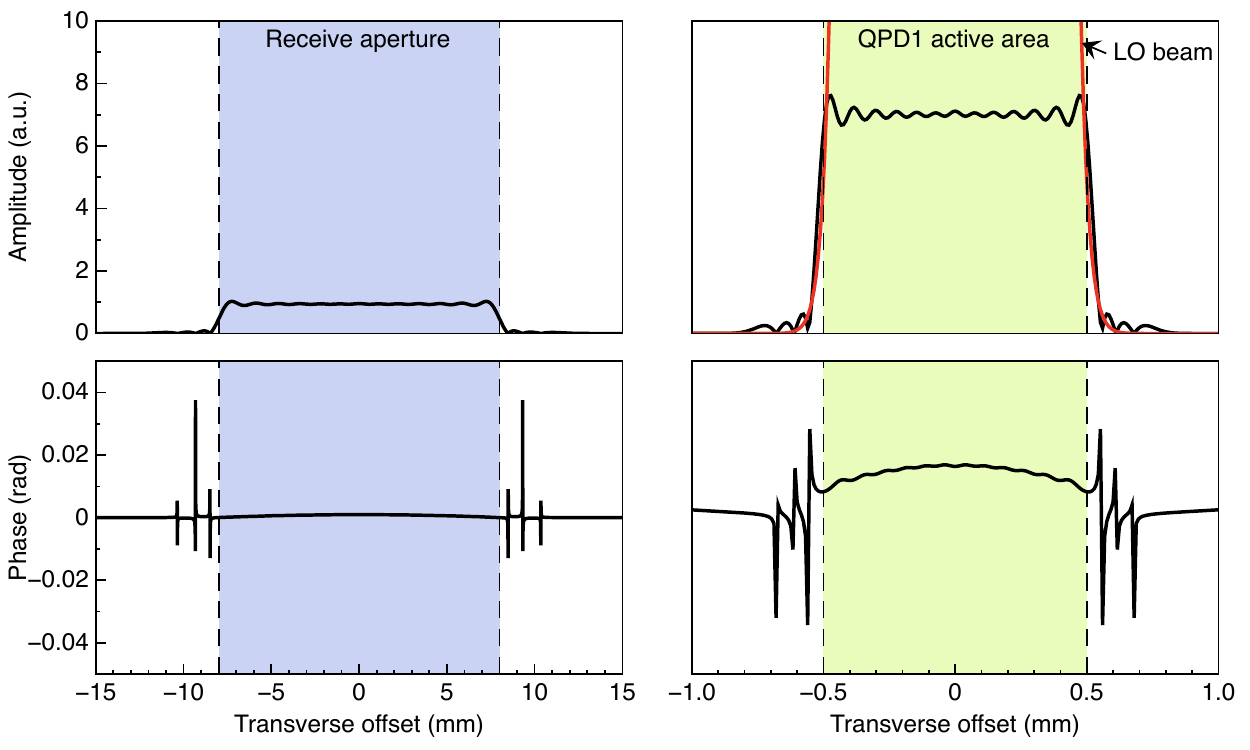}
\caption{Amplitude and phase of the ``flat-top'' RX beam at the receive aperture and at the surface of QPD1, in the horizontal direction. The dashed lines indicate the boundaries of the aperture and the active area of the photodiode respectively. The RX beam is modelled in IFOCAD using the mode-expansion method. The amplitude of the local oscillator beam at QPD1 is shown in red, showing good spatial overlap with the RX beam, in spite of the much larger peak amplitude.}
\label{figure:mem}
\end{figure}

As shown in Figure~\ref{figure:setup}, the RX beam, which has propagated for the inter-satellite distance of $\sim 200\,$km, is clipped to a radius of 8\,mm by the RX/TX RP aperture located on the OB, and the diffracted beam's amplitude and phase profile is nearly flat. To simulate this ``flat-top'' beam, the mode-expansion method (MEM)~\cite{Mahrdt2014PhD} is adopted. In the MEM, the electrical field of the incoming Gaussian beam $E(x,y)$ is decomposed into Hermite-Gaussian modes $u_{mn}(x,y;q)$,
\begin{equation}
E_{\rm MEM}(x,y)=\sum^{N_{\rm max}}_{m=0}\sum^{N_{\rm max}-m}_{n=0}a_{mn}u_{mn}(x,y;q)\exp(-iks), \label{eq:mem}
\end{equation}
where $a_{mn}$ is the mode amplitude, $k$ is the wavenumber, $s$ is a propagation distance, and $N_{\rm max}$ is the maximum expansion order, which limits the decomposition performance. Once the relative mode amplitudes are derived at the aperture, they are invariant through the paraxial propagation of the modes to the detectors downstream. Hence, the electrical field at an arbitrary plane along the optical axis can be computed as the sum over the Hermite-Gaussian modes propagated up to that plane. Figure~\ref{figure:mem} shows the amplitude and phase of the RX beam at the receive aperture and at the surface of QPD1, as obtained via the MEM simulation.

\begin{figure}[h]
\includegraphics[scale=1]{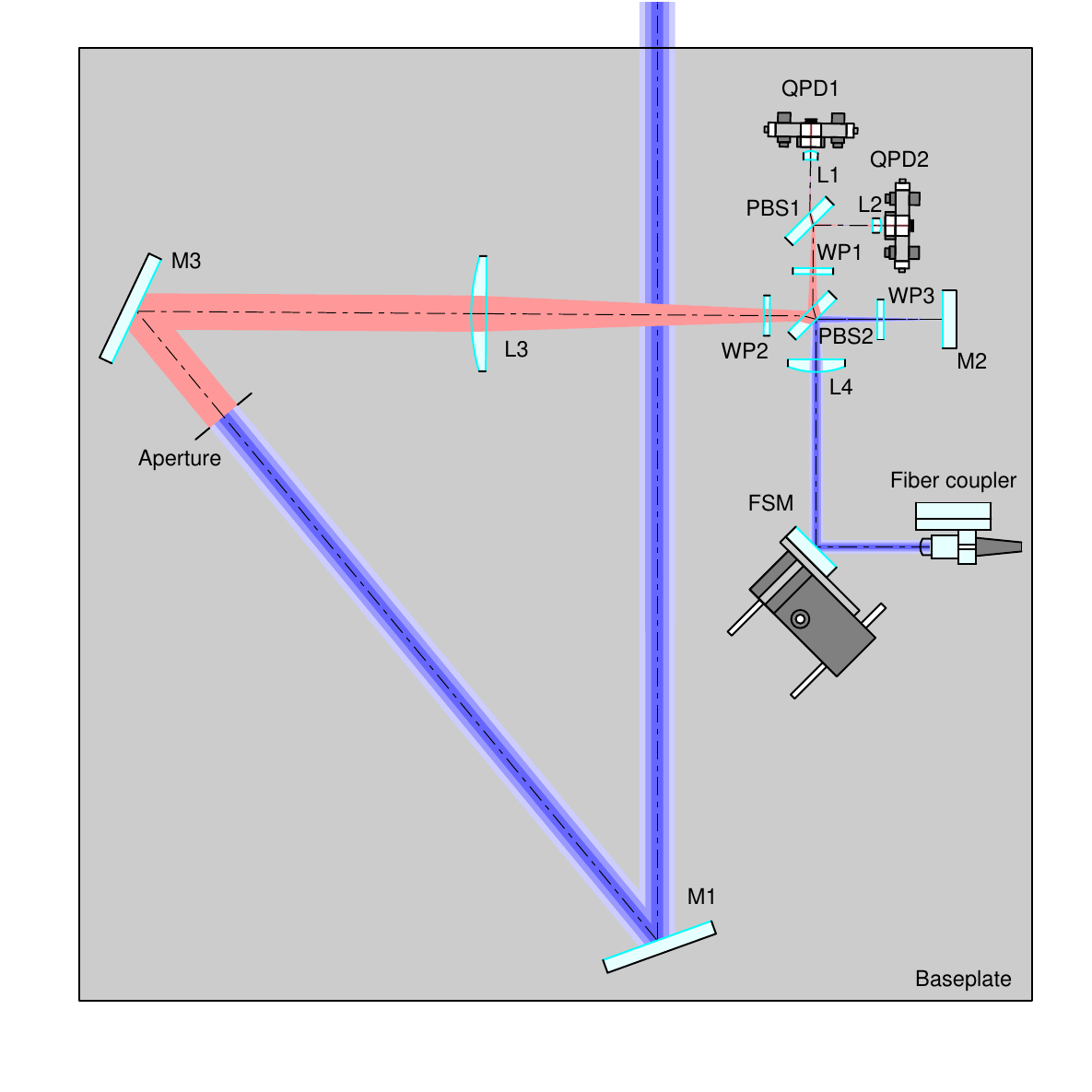}
\caption{LRI OB as modelled in IFOCAD and drawn in OPTOCAD. The RX beams starts at the receive aperture (red). The TX and LO beams stem from the beam injected into the OB by the fiber injector (blue). The baseplate assumed in the thermal analysis is drawn as a rectangle enclosing all of the OB components.}
\label{figure:ob-sim}
\end{figure}

The IFOCAD model is built based on the ``B'' configuration from Table~\ref{table:configurations}, and it is drawn using OPTOCAD, a Fortran 95 module for tracing Gaussian TEM00 beams through an optical setup~\cite{Optocad} (Figure~\ref{figure:ob-sim}). The B configuration is chosen due to the smaller size of the beam injected into the OB, which is straightforward to realize using commercially available fiber injectors.

 After the OB components are initialized in the model, the imaging systems are optimized by tuning the positions of the photodiodes, of lens L3, of the receiving aperture, and of mirror M2, in the following steps: first, the QPD positions are fine-tuned along the nominal optical axis in order to minimize the beam walk of the LO beam on the detector planes, under both pitch and yaw motions of the steering mirror; second, the position of L3 is fine-tuned to minimize the separation between the front focal point of L1 and the effective focal point of the system composed of L3, WP2, PBS2, WP1 and PBS1; third, the position of the receiving aperture is fine-tuned to minimize the interferometer's TTL coupling, as sensed by both quadrant photodiodes, under pitch and yaw rotations of the S/C with an active steering mirror (i.e., using the DWS signals to actuate on the steering mirror's pitch and yaw degrees of freedom in order to yield optimally overlapped phasefronts between the RX and LO beams at the detectors); lastly, the position of M2 is fine-tuned such that the effective focal point of the TX IS is located at the RX/TX RP, and thus near-perfect TX beam pointing to the remote S/C is achieved. In the resulting configuration, the interferometer with closed-loop control of the steering mirror becomes largely insensitive to the angular motion of the S/C.

\begin{figure}
\includegraphics[scale=0.15]{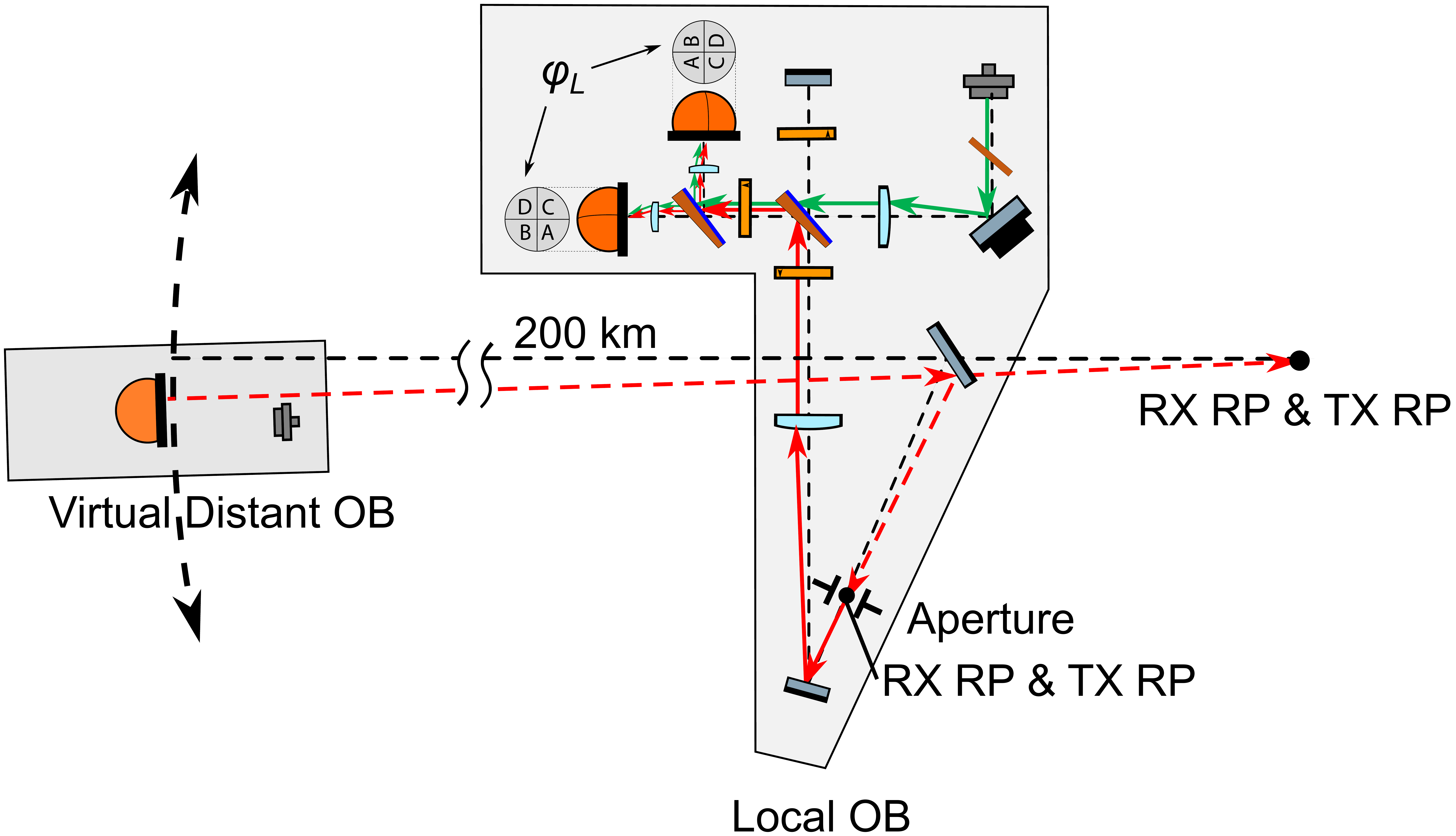}
\caption{Depiction of local S/C TTL coupling simulation. The RX beam (red) starts at the local RX RP with the expected size after having propagated for the inter-satellite distance, and a certain rotation angle to simulate S/C angular motion. The RX beam interferes with the LO beam (green) and their beatnote is captured by the photodetectors.}
\label{figure:local-ttl}
\end{figure}

\begin{figure}
\includegraphics[scale=0.15]{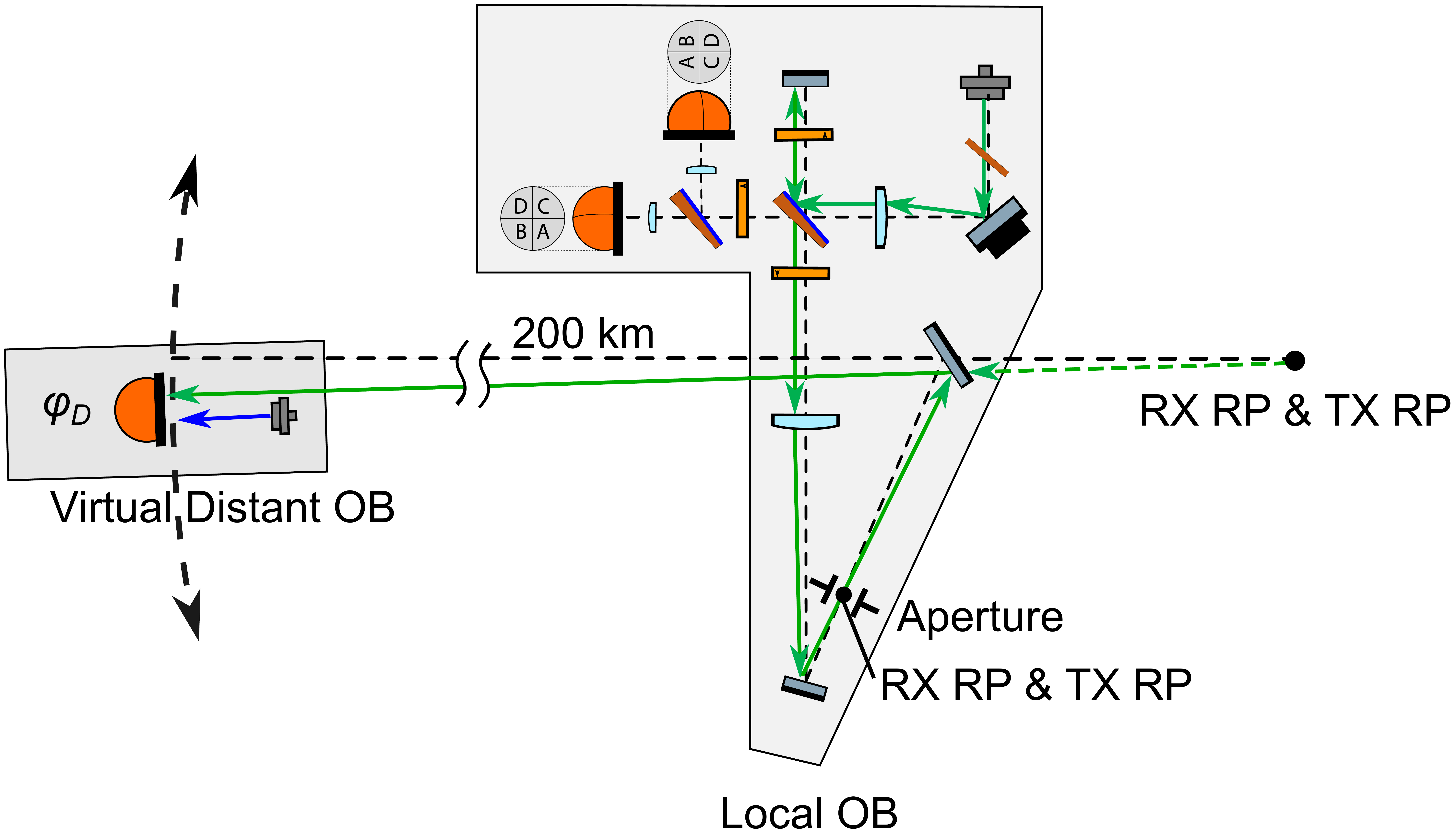}
\caption{Depiction of the TX beam TTL coupling simulation. The TX beam (green) propagates to the preset distant OB where it interferes at a single-element photodiode with a Gaussian beam large enough such that the distant system acts as a perfect transponder.}
\label{figure:distant-ttl}
\end{figure}

These functionalities are quantitatively evaluated at the local S/C, and TTL coupling is measured both at the local S/C and at the distant S/C in perfect-transponder mode. First, an RX beam of the expected size after having propagated through the inter-satellite path is originated at the local RX RP, i.e., at the center of the receive aperture (Figure~\ref{figure:local-ttl}). This beam is tilted in the pitch and yaw degrees of freedom to simulate angular motion of the S/C around its c.m.; The RX beam is propagated through the setup, and it is interfered with the LO beam whilst the FSM is being actively controlled using DWS loops to obtain optimal RX-LO overlap at the detectors. The LPS variations are measured, and the local TTL coupling is evaluated by varying the misalignment angles.

Then, the position of a distant single-element photodiode (SEPD) representing the remote OB is derived by intersecting the initial RX beam direction with a sphere of 200\,km radius centered around the unfolded RX RP (Figure~\ref{figure:distant-ttl}). The locally generated TX beam is propagated to this SEPD in the far field, where it is interfered with a large Gaussian beam acting as LO beam for the remote S/C OB. The radius of this LO beam is set large enough such that this distant system acts as a perfect transponder, i.e., such that it measures a range unaffected by the wavefront of the interfering beams. The LPS measured by the distant SEPD is obtained and the TX beam TTL coupling is evaluated.

\begin{figure}
    \centering
    \includegraphics{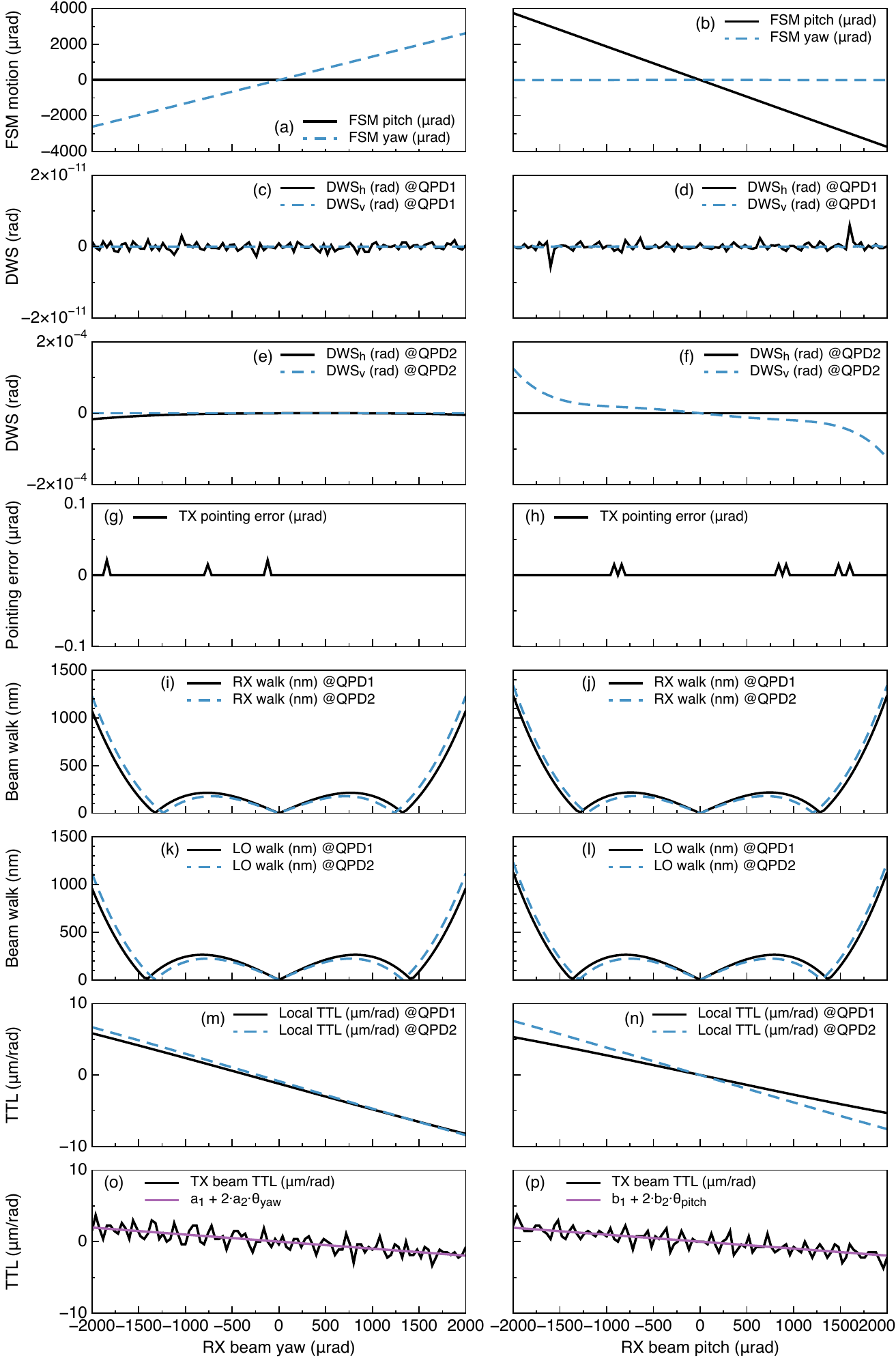}
    \caption{Angular motion of the local S/C causes the received beam's incidence angle at the receive aperture to change (the abscissas). A steering mirror is actuated (a, b) via two independent DWS control loops that zero the $\text{DWS}_{\text{h}}$ and $\text{DWS}_{\text{v}}$ signals in QPD1 (c, d), keeping the RX and LO phasefronts nearly parallel at both detectors. QPD2 is out of loop and measures nearly zero $\text{DWS}_{\text{h}}$ and $\text{DWS}_{\text{v}}$ (e, f). The TX IS ensures that the TX beam is antiparallel to the RX beam in the inter-S/C path (g, h). The RX IS and LO IS ensure that the RX and LO beams experience minimal beam walk at the detectors (i-l). In the resulting configuration, TTL coupling is minimized, as measured both in the local S/C (m, n), and at the distant S/C (o, p).}
    \label{figure:opt-sim-1}	
\end{figure}

The results are depicted in Figure~\ref{figure:opt-sim-1}, which shows the variation of relevant quantities as a function of RX beam angle in both the pitch and yaw degrees of freedom for the interferometer with closed-loop control of the steering mirror. The figure shows (in order from top to bottom) the position acquired by the active steering mirror under a given misalignment; the measured in-loop (QPD1) and out-of-loop (QPD2) horizontal and vertical DWS signals; the deviation of the transmitted beam with respect to the inter-satellite line of sight; the RX and LO beam walk at the photodetectors; the TTL coupling experienced at the local S/C; and the TX beam TTL coupling measured at the distant S/C in perfect-transponder mode. Note that the out-of-loop DWS signals are small but orders of magnitude greater than the in-loop signals due to the asymmetry between the reflection and transmission ports of PBS1. The TX beam TTL coupling is limited by the numerical error of double precision. 

The installation error of M2 directly affects the OB's retro-reflective function by changing the position of the TX RP, thereby introducing extra TTL coupling to the range measurement. This effect is modelled and characterized, and it is determined that with positional tolerances of 0.05\,mm and angular tolerances of 50\,${\upmu}$rad, the extra TTL coupling in all three rotation axes is $\sim 10$\,${\upmu}$m/rad. This performance impact is similar to that of the triple-mirror-assembly of the GRACE-FO LRI, where the misalignment of the three mirrors introduced a TTL coupling of less than 20\,${\upmu}$m/rad in all three rotation axes\cite{schutze2015PhD}.

Finally, a thermal analysis is carried out to ensure that the proposed OB layout is robust against thermoelastic deformation and refractive index variations due to temperature fluctuations. This analysis is of particular importance due to the fact that the RX and LO imaging systems both feature a different set of components on their respective sensitive paths. This means that, e.g., any thermal fluctuations introduced by lenses L3 and L4 are not common-mode between the interfering beams.

\begin{figure}
    \centering
    \includegraphics[scale=1.0]{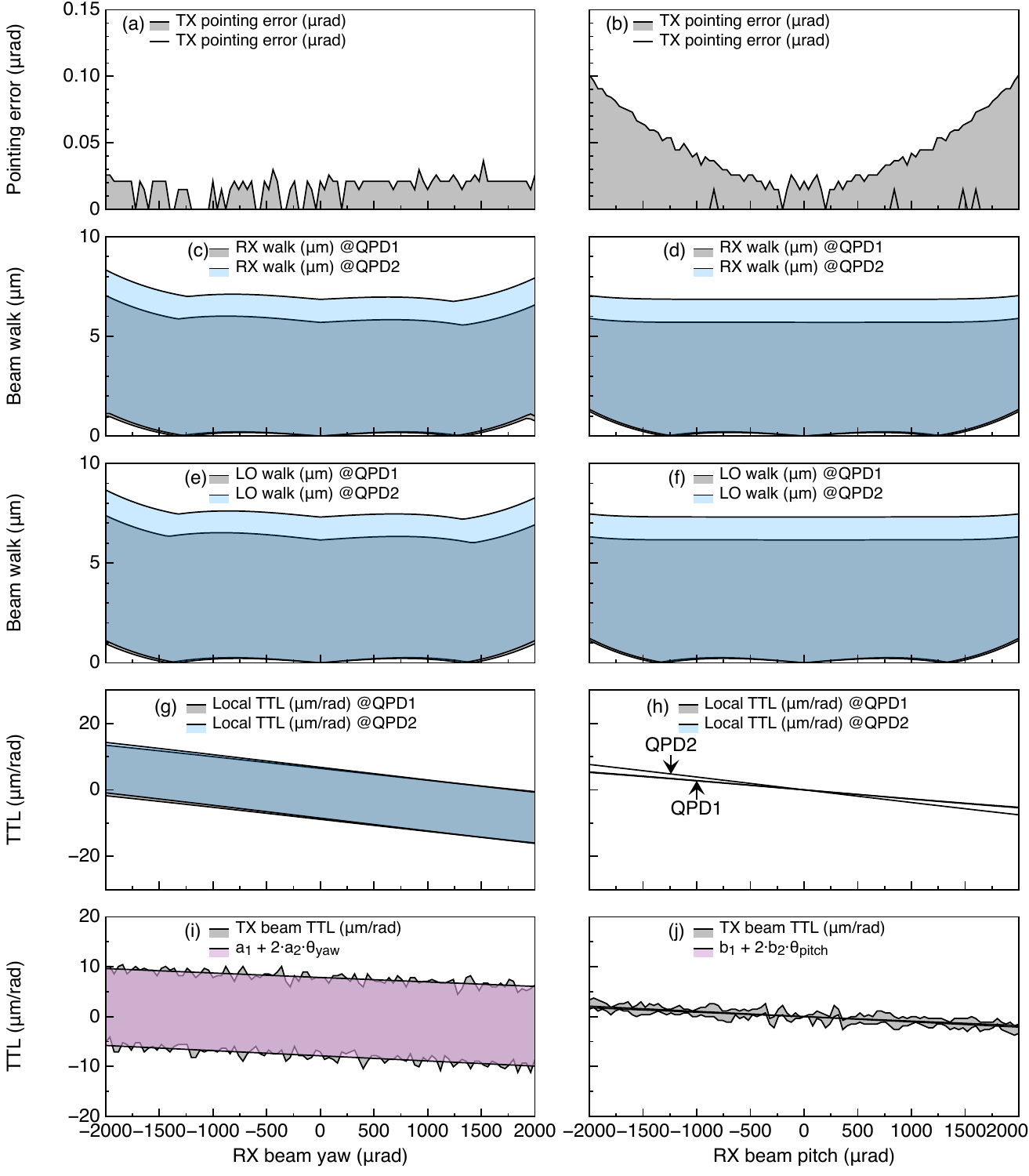}
    \caption{Thermal analysis. The temperature is swept in the $\pm 3$\,K range in seven steps, and the TTL coupling simulation is carried out for each step of temperature and RX beam tilt in the pitch and yaw degrees of freedom. The shaded regions shown in the plots are bounded by the maximum and minimum results for TX pointing error (a, b), RX beam walk (c, d), LO beam walk (e, f), local TTL coupling (g, h), and	 TX beam TTL coupling (i, j) throughout the temperature range.}
    \label{figure:thermal_analysis}
\end{figure}

The expected thermal drift on the optical bench is $\pm 3$\,K/orbit~\cite{Kornfeld2019}, hence, the temperature is swept within this range and both the thermoelastic expansion and the change of refractive index are coherently applied to all optical components. The components are assumed to be attached to a baseplate that expands uniformly, changing the components' positions w.r.t.~the baseplate's c.m.; For the optical components, the thermal coefficients of fused silica are used, $\alpha=5.5\times10^{-7}$\,1/K, and $\beta=9.6\times10^{-6}$\,1/K, where $\alpha$ measures the fractional length change ($\Delta L / L$ per Kelvin) and $\beta$ measures the fractional refractive index change ($\Delta n / n$ per Kelvin). On the other hand, the baseplate is assumed to be made of Titanium with $\alpha=8.6\times 10^{-6}$\,1/K. The results are shown in Figure~\ref{figure:thermal_analysis}. This shows the maximum magnitude of the total TTL coupling of $18.1$\,$\upmu$m/rad and $1.16$\,$\upmu$m/rad over $300$\,$\upmu$rad in yaw and pitch, respectively. These are both below the GRACE-FO requirement of $80$\,$\upmu$m/rad~\cite{Nicklaus2017}, and thus verify the feasibility of the design in terms of the orbital thermal drift.

\section{Summary and outlook}
\label{section:conclusions}

Following on the success of the GRACE-FO LRI, a US-German collaboration is conducting preparatory studies for advancing the LRI design and technology under the GRACE-ICARUS project~\cite{Flechtner2020}. In parallel, ESA has approved preparatory studies and instrument pre-developments for a Next Generation Gravity Mission~\cite{Nicklaus2020, Haagmans2020}. In this context, we have put forward an optical bench design for a next-generation LRI.

The design follows an on-axis topology in which the RX, TX and LO beams' paths coincide. A series of Keplerian telescope imaging systems are configured to: $i.$ achieve the retro-reflective function without the need for a retroreflector; $ii.$ adapt the size of the received beam and the local oscillator beam to the active area of the quadrant photodiodes; and $iii.$ adjust the size of the transmitted beam to match the size of the RX/TX aperture and decrease its divergence to relax the laser power requirements. 

The imaging systems, in combination with an active steering mirror using differential wavefront sensing loops, ensure optimal overlap between the interfering RX and LO beams at the detectors, as well as near-perfect TX beam pointing to the remote S/C. The optimal RX/LO overlap maximizes the local carrier-to-noise ratio, and the accurate TX beam pointing maximizes the optical power at the remote S/C, regardless of the attitude of the local S/C. 

The optical bench functionalities and limitations are demonstrated by means of a computer model using the IFOCAD C++ libraries. The coupling of the attitude of the local S/C to the inter-satellite range measurement is computed, and it is found to be well within the required level of $100\,\upmu$m/rad. A thermal analysis accounting for the effects of thermoelastic expansion of the optical bench components and baseplate, and refractive index fluctuations of the optics, is carried out to verify the robustness of the proposed design against the temperature fluctuations expected in orbit.

The proposed layout uses a single telescope in the RX and TX beam path, enhancing the light collecting area and decreasing the TX beam far field divergence. Since this design requires only a single baffle, the footprint of the LRI in the S/C can be made smaller compared to off-axis designs like the GRACE-FO LRI, which hosted a retroreflector that was separate from the OB assembly, leading to significantly increased complexity of assembly and thermal control.

An experimental realization of the proposed OB design is being conducted at the Albert Einstein Institute. While the primary functions of the OB have been demonstrated via the optical simulations presented in this paper, several questions remain to be investigated experimentally, notably the coupling of polarization imperfections and fluctuations, and stray light, into the length and angular measurements.

\vspace{6pt}

\authorcontributions{Conceptualization, Y.Y. and V.M.; methodology, K.Y. and M.D.A; software, K.Y., M.D.A., and D.W.; resources, J.J.E.D.; writing---original draft preparation, Y.Y., K.Y., M.D.A, D.W., J.J.E.D. and V.M.; writing---review and editing, M.D.A.; project administration, M.D.A., J.J.E.D., V.M., J.J., and G.H.; funding acquisition, G.H.; All authors have read and agreed to the published version of the manuscript.}

\acknowledgments{This work has been supported by: the Chinese Academy of Sciences (CAS) and the Max Planck Society (MPG) in the framework of the LEGACY cooperation on low-frequency gravitational-wave astronomy (M.IF.A.QOP18098); The Deutsche Forschungsgemeinschaft (DFG, German Research Foundation, Project-ID 434617780, SFB 1464). Clusters of Excellence “QuantumFrontiers: Light and Matter at the Quantum Frontier: Foundations and Applications in Metrology” (EXC-2123, project number: 390837967); PhoenixD: “Photonics, Optics, and Engineering – Innovation Across Disciplines” (EXC-2122, project number: 322 390833453).}

\conflictsofinterest{The authors declare no conflict of interest.}

\abbreviations{The following abbreviations are used in this manuscript:\\

\noindent
\begin{tabular}{ll}
c.m. & Center-of-mass \\
CNR & Carrier to noise density \\
DLR & Deutsches Zentrum f\"ur Luft- und Raumfahrt \\
DWS & Differential Wavefront Sensing \\
ESA & European Space Agency \\
FSM & Fast steering mirror \\
GPS & Global Positioning System \\
GRACE & Gravity Recovery and Climate Experiment \\
GRACE-FO & Gravity Recovery and Climate Experiment Follow-On \\
IS & Imaging system \\
LISA & Laser Interferometer Space Antenna \\
LO & Local oscillator (beam) \\
LPS & Longitudinal pathlength signal \\
LRI & Laser Ranging Interferometer \\
MEM & Multimode expansion method \\
NASA & National Aeronautics and Space Administration \\
OB & Optical Bench \\
PBS & Polarizing beamsplitter \\
QPD & Quadrant Photodiode \\
RP & Reference point \\
RX & Received (beam) \\
S/C & Spacecraft \\
SEPD & Single-element photodiode \\
TTL & Tilt-to-length (coupling) \\
TX & Transmit (beam) \\
\end{tabular}}

\vspace{12pt}

\noindent
\textbf{List of commonly used symbols}

\noindent
\begin{tabular}{ll}
$m_i~\mathrm{with}~i \in \left\{\mathrm{RX,LO,TX} \right\} $ & Angular magnification of the $i$-beam \\
$\alpha_{i,\mathrm{IN}}~\mathrm{with}~i \in \left\{\mathrm{RX,LO,TX} \right\} $ & Angle of the $i$-beam at the entrance pupil \\
$\alpha_{i,\mathrm{OUT}}~\mathrm{with}~i \in \left\{\mathrm{RX,LO,TX} \right\} $ & Angle of the $i$-beam at the exit pupil \\
$r_{\mathrm{TX/RX AP}}$ & Receive and transmit beam aperture radius \\
$\omega_0$ & Waist radius of the LO beam at the fiber injector \\
$\eta$ & Heterodyne efficiency \\
$\theta_{\mathrm{TX}}$ & Half-angle divergence ($1/e^2$) of the TX beam \\
$\phi$ & Interferometric phase measured by a SEPD \\
$\phi_{A...D}$ & Interferometric phase measured by a segment of a QPD \\
$ k = 2 \pi / \lambda$ & The wavenumber, with $\lambda$ the wavelength \\
\end{tabular}

\end{paracol}
\reftitle{References}

%\externalbibliography{yes}
%\bibliography{ref}

\end{document}